\documentclass{hep99}
\usepackage{graphicx}
\begin{document}

\title{\vskip-1.5cm{\large\hfill LU TP 99--29\\\hfill October 1999\\}
The $\Delta I=1/2$ rule and other matrix elements$^{\dag\ddag}$}

\author{J. Bijnens$^1$}
%

\address{$^1$ Department of Theoretical Physics 2, Lund University\\
S\"olvegatan 14A, S22362 Lund, Sweden\\[3pt]
E-mail: {\tt bijnens@thep.lu.se}}

\abstract{Recent work by J.Prades and myself on $K\to\pi\pi$ is described.
The method we use to consistently connect long and short distances
is described and numerical results for the $\Delta I=1/2$ rule and on
$B_6$, the parameter relevant for the strong part of $\epsilon'/\epsilon$,
are given in the chiral limit.
} 

\maketitle

\fntext{\dag}{Work
supported in part by TMR, EC-Contract No. ERBFMRX-CT980169
(EURO\-DA$\Phi$NE).}
\fntext{\ddag}{Talk presented at EPS HEP99, Tampere, Finland, July 15-21,1999}

\section{Introduction}
The qualitative feature that
$\Gamma(K^0\to\pi^0\pi^0) \gg \Gamma(K^+\to\pi^+\pi^0)$
is one of the oldest problems in kaon physics, the $\Delta I=1/2$ rule.
The isospin-2 final state amplitude $A_2$ is much smaller
than the isospin-0 amplitude $A_0$, experimentally
$|A_0/A_2| = 22.1$, while simple $W$-exchange naively predicts a ratio
of $\sqrt{2}$. The work presented here has been published in \cite{BP1}
and presented in \cite{BPtalks}.
A review of Kaon physics is in \cite{kreview} and in the talks presented
at Kaon99\cite{Kaon99}.

The underlying standard model process is the exchange of a $W$-boson
but the large difference in the Kaon and $W$-mass enhances
normally suppressed contributions by large factors
$\ln(m_W^2/m_K^2)\approx 10$. At the same time, at low energies the strong
interaction coupling $\alpha_S$ becomes very large which requires us to use
non-perturbative methods at those scales.

The resummation of large logarithms at short-distance can be done
using renormalization group methods. At a high scale the exchange
of $W$-bosons is replaced by a sum over local operators. For weak decays
these start at dimension 6. The scale can then be lowered using the
renormalization group.
The short-distance running is now known to two-loops 
\cite{two-loops1,two-loops2} (NLO)
which sums the $\left(\alpha_S\ln(m_W/\mu)\right)^n$ and
$\alpha_S\left(\alpha_S\ln(m_W/\mu)\right)^n$ terms. A review of this can
be found in the lectures by A. Buras \cite{Buras}.

The major remaining problem is to calculate the matrix elements
of the local operators at some low scale. I will address some progress
on this issue in this talk. The main method was originally
proposed in Ref. \cite{BBG} arguing that $1/N_c$ counting could
be used to systematically calculate the matrix elements.
Various improvements have since been
introduced. The correct momentum routing was introduced
in \cite{BBG2}. The use of the extended Nambu-Jona-Lasinio model as
an improved low energy model was introduced for weak matrix
elements in \cite{BP2} and a short discussion of its major
advantages and disadvantages can be found in \cite{BPP}.
The results obtained were encouraging but a major problem remained.
At NLO order the short-distance running becomes dependent
on the precise definition of the local operators. This dependence
should also be reflected in the calculations of the matrix elements
as well as a correct identification of the scale of the
renormalization group in the matrix element calculation.
The more precise interpretation of the scheme of \cite{BBG}
introduced in \cite{BP2} was shown there at one-loop to satisfy the
latter criterion. I present in the next section
how this method also satisfies the latter
at NLO and how it solves the first problem as well. We call
this method the $X$-boson method. The third section describes the
numerical results
we obtained in \cite{BP1} for the $\Delta I=1/2$ rule in the chiral limit.
The results obtained there are also reported here in the more
standard $B_6$, defined here with respect to our X-boson scheme.

Other recent work on matrix elements is the work of
\cite{Hambye} and \cite{eduardo} using the $1/N_c$ method as well.
A more model dependent approach is \cite{Trieste}.

\section{The $X$-boson method}
The basic idea is that we know how to hadronize
currents or at least that this is a tractable problem. So we replace
the effect of the local operators of
$H_W(\mu) = \sum_i C_i(\mu) Q_i(\mu)$ at a scale $\mu$
by the exchange of a series of colourless $X$-bosons at a low scale $\mu$.
The scale $\mu$ should be such that the $1/N_c$ suppressed contributions
have no longer large logarithmic corrections.
Let me illustrate the procedure in the case of only one operator
and neglecting penguin contributions.
In the more general case all coefficients become matrices.
\begin{eqnarray}
\lefteqn{C_1(\mu)(\bar s_L\gamma_\mu d_L)(\bar u_L\gamma^\mu u_L)
\Longleftrightarrow}&&\nonumber\\&&
X_\mu\left[g_1 (\bar s_L\gamma^\mu d_L)+g_2 (\bar u_L\gamma^\mu u_L)
\right]\,.
\end{eqnarray}
Summation over colour indices inside brackets is understood.
We now determine $g_1$, $g_2$ as a function of $C_1$. This is done by
equalizing matrix elements of $C_1 Q_1$ with the equivalent ones
of $X$-boson exchange. The matrix elements are at the scale $\mu$ chosen
such that perturbative QCD methods can still be used and thus we can use
external states of quarks and gluons.
\begin{figure}[htb]
\begin{center}
\includegraphics[width=0.8\columnwidth]{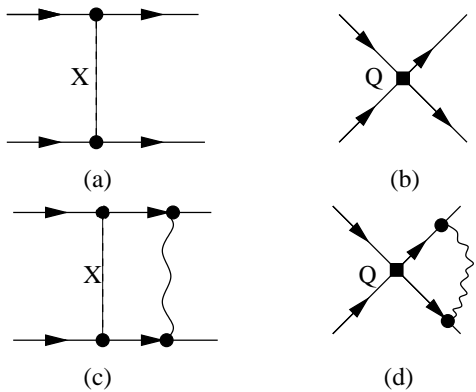}
\caption{\label{figX} The diagrams needed for the identification
of the local operator $Q$ with $X$-boson exchange in the case of
only one operator and no Penguin diagrams. The wiggly line
denotes gluons, the square the operator Q and the dashed line
the $X$-exchange. The external lines are quarks.}
\end{center}
\end{figure}
To lowest order this is simple. The tree level diagram
from Fig. \ref{figX}(a) is set equal to that of Fig. \ref{figX}(b)
leading to 
\begin{equation}
C_1 = {g_1 g_2}/{M_X^2}\,.
\end{equation}
At NLO diagrams 
like those of Fig. \ref{figX}(c)
and \ref{figX}(d) contribute as well leading to
\begin{eqnarray}
\lefteqn{C_1\left(1+\alpha_S(\mu)r_1\right)
=}&&\nonumber\\&& \frac{g_1 g_2}{M_X^2}\left(1+\alpha_S(\mu)a_1+
\alpha_S(\mu)b_1\log\frac{M_X^2}{\mu^2}\right)\,.
\end{eqnarray}
At this level the scheme-dependence disappears. The left-hand-side (lhs)
is scheme-independent. The right-hand-side can be calculated in a
very different renormalization scheme from the lhs.
The infrared dependence of $r_1$ is
present in  precisely the same
way in $a_1$ such that $g_1$ and $g_2$ are scheme-independent
and independent of the precise infrared definition of the external state
in Fig. \ref{figX}.

One step remains, we now have to calculate the matrix element
of $X$-boson exchange between meson external states.
The integral over $X$-boson momenta we split in two
\begin{eqnarray}
\label{split}
\lefteqn{
\int_0^\infty dp_X\frac{1}{p_X^2-M_X^2}
\Longrightarrow}&&\nonumber\\&&
\int_0^{\mu_1}dp_X\frac{1}{p_X^2-M_X^2}
+\int_{\mu_1}^\infty dp_X\frac{1}{p_X^2-M_X^2}\,.
\end{eqnarray}
The second term involves a high momentum that needs to flow back
through quarks or gluons and leads through diagrams like the one
of Fig. \ref{figX}(c)
to a four quark-operator with a coefficient
\begin{equation}
\frac{g_1 g_2}{M_X^2}\left(\alpha_S(\mu_1)a_2
+\alpha_S(\mu_1)b_1\log\frac{M_X^2}{\mu^2}\right)\,.
\end{equation}
The four-quark operator thus
needs to be evaluated only in leading order in $1/N_c$.
The first term we have to evaluate in a low-energy model with as much
QCD input as possible.
The $\mu_1$ dependence cancels between the two terms in (\ref{split})
if the low-energy model is good enough and all dependence on
$M_X^2$ cancels out to the order required as well.
Calculating the coefficients $r_1$, $a_1$ and $a_2$ gives the
required correction to the naive factorization method as used
in previous $1/N_c$ calculations.

It should be stressed that in the end all dependence on $M_X$ cancels
out. The $X$-boson is a purely technical device to correctly
identify the four-quark operators in terms of well-defined products of
nonlocal currents.

\section{Numerical results and conclusions}
We now use the $X$-boson method with $r_1$ as given in \cite{two-loops1}
and $a_1=a_2=0$, the calculation of the latter are in progress,
and $\mu=\mu_1$. For $B_K$ we can extrapolate to the pole
for the real case ($\hat B_K$) and in the chiral limit ($\hat B_K^\chi$)
and for $K\to\pi\pi$ we can get at the
values of the octet ($G_8$), weak mass term ($G_8^\prime$)
and 27-plet ($G_{27}$) coupling.
We obtain
\begin{displaymath}
\hat B_K = 0.69\pm0.10\,;~ \hat B_K^\chi= 0.25\mbox{--}0.40\,;~
G_8= 4.3\mbox{--}7.5\,;
\end{displaymath}
\begin{equation}
G_{27}=0.25\mbox{--}0.40\mbox{ and }
G_8^\prime=0.8\mbox{--}1.1\,,
\end{equation}
to be compared with the experimental values
$G_8\approx6.2$ and $G_{27}\approx0.48$ \cite{BP1,Kambor}.

\begin{figure}
\includegraphics[width=\columnwidth]{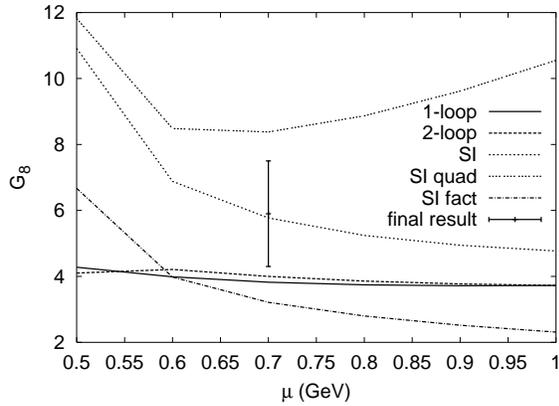}
\caption{\label{figg8} The octet coefficient $G_8$ as a function of
$\mu$ using the ENJL model and the one-loop Wilson coefficients,
the 2-loop ones and those including the $r_1$ (SI). In
the latter case also the factorization (SI fact)
and the approach of \cite{Hambye} (SI~quad) are shown.}
\end{figure}

\begin{figure}
\includegraphics[width=\columnwidth]{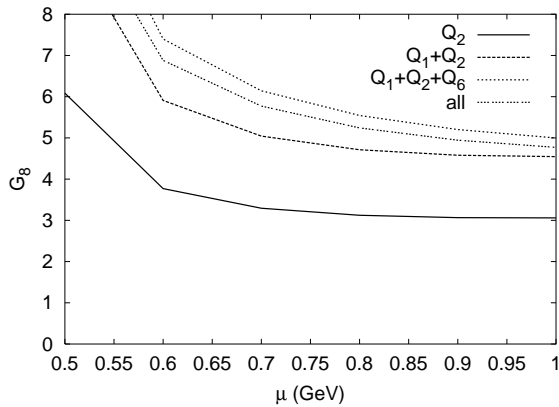}
\caption{\label{figg8_comp} The composition of $G_8$ as a function of
$\mu$. Shown are $Q_2$, $Q_1+Q_2$, $Q_1+Q_2+Q_6$ and all 6 $Q_i$.
The coefficients $r_1$ are included in the Wilson coefficients.}
 \end{figure}

In Fig. \ref{figg8} the $\mu$ dependence of $G_8$ is shown
and in Fig. \ref{figg8_comp} the contribution from the various
different operators.

Fig. \ref{figg8_comp} shows that the contribution from $Q_6$ to
the $\Delta I=1/2$ rule is rather small. It is the penguin like
contributions from $Q_2$ that are the major contributions.
From our numerics we can also extract the value from $B_6$ that follows
from our calculation.
In Table \ref{tab:B6} we give the value as a function of
the matching scale $\mu$ for the calulation in CHPT and the one using
the ENJL model. We have normalized here to the large $N_c$ value since the pure
factorizable value of $B_6$ is ill-defined in the chiral limit\cite{BP1}.
The enhancement away from lower values was also seen in the most recent
paper of \cite{Hambye}.

\begin{table}
\begin{center}
\caption{\label{tab:B6} $B_6$ as a function of $\mu$ using CHPT and the ENJL
model. Numbers are calculated using the results of \cite{BP1}.}
\begin{tabular}{c|ccccc} 
\br
$\mu$ (GeV) & 0.6 & 0.7 & 0.8 & 0.9 & 1.0 \\
\mr
CHPT & 1.19 & 0.93 & 0.70 & 0.50 & 0.36 \\
ENJL & 2.27 &2.16 & 2.11&  2.11& 2.14\\
\br
\end{tabular}
\end{center}
\end{table}

I showed how the $X$-boson method allows to correctly treat NLO
scheme dependence and that using that method and the
ENJL model at low energies reproduces the $\Delta I=1/2$ rule
{\em quantitatively} without any free parameters.
The results for $B_6$ are encouraging with respect to the experimental
value of $\epsilon'/\epsilon$

\end{document}